\documentclass[sigconf]{acmart}
\usepackage[]{float}
\usepackage[]{multirow}
\usepackage[]{color}
\usepackage{subcaption}
\usepackage{hyperref}

\AtBeginDocument{%
  }

\setcopyright{acmcopyright}
\copyrightyear{2018}
\acmYear{2018}
\acmDOI{XXXXXXX.XXXXXXX}

\acmConference[Conference acronym 'XX]{Make sure to enter the correct
  conference title from your rights confirmation emai}{June 03--05,
  2018}{Woodstock, NY}
\acmPrice{15.00}
\acmISBN{978-1-4503-XXXX-X/18/06}

\begin{document}

\title{ControlRec: Bridging the Semantic Gap between Language Model and Personalized Recommendation}


\author{Junyan Qiu}
\email{qiujunyan2018@ia.ac.cn}
\orcid{0000-0001-9316-8213}
\affiliation{%
  \institution{University of Chinese Academy of Sciences}
  \city{Beijing}
  \country{China}
}

\author{Haitao Wang}
\email{wanghaitao13@meituan.com}
\affiliation{%
  \institution{Meituan}
  \city{Beijing}
  \country{China}
}
\author{Zhaolin Hong}
\affiliation{%
  \institution{Meituan}
  \city{Beijing}
  \country{China}}
\email{hongzhaolin@meituan.com}

\author{Yiping Yang}
\affiliation{%
  \institution{Institute of Automation, Chinese Academy of Sciences}
  \city{Beijing}
  \country{China}}
\email{yiping.yang@ia.ac.cn}

\author{Qiang Liu}
\affiliation{%
  \institution{Meituan}
  \city{Beijing}
  \country{China}}
\email{liuqiang43@meituan.com}

\author{Xingxing Wang}
\affiliation{%
  \institution{Meituan}
  \city{Beijing}
  \country{China}}
\email{wangxingxing04@meituan.com}
%
%
%
%
%
%


\begin{abstract}
	The successful integration of large language models (LLMs) into recommendation systems has proven to be a major breakthrough in recent studies, paving the way for more generic and transferable recommendations. However, LLMs struggle to effectively utilize user and item IDs, which are crucial identifiers for successful recommendations. This is mainly due to their distinct representation in a semantic space that is different from the natural language (NL) typically used to train LLMs. To tackle such issue, we introduce \textbf{ControlRec}, an innovative \textbf{Contr}astive pr\textbf{o}mpt \textbf{l}earning framework for \textbf{Rec}ommendation systems. ControlRec treats user IDs and NL as heterogeneous features and encodes them individually. To promote greater alignment and integration between them in the semantic space, we have devised two auxiliary contrastive objectives: (1) Heterogeneous Feature Matching (HFM) aligning item description with the corresponding ID or user's next preferred ID based on their interaction sequence, and (2) Instruction Contrastive Learning (ICL) effectively merging these two crucial data sources by contrasting probability distributions of output sequences generated by diverse tasks. Experimental results on four public real-world datasets demonstrate the effectiveness of the proposed method on improving model performance. 
\end{abstract}




\maketitle

\section{Introduction}
Recommendation systems are designed to tailor recommendations to individual users based on their unique interests and preferences \cite{cheng2016wide,guo2017deepfm,zhou2018deep}. Such systems often have to balance multiple objectives simultaneously \cite{ma2018modeling,tang2020progressive,deng2023unified}. For example, they need to predict a user's preferences for a movie while also offering explanations for why the user is inclined towards the particular movie. Given that recommendation tasks normally involve a shared user-item pool and overlapping contextual features \cite{geng2022recommendation}, it is crucial to develop a single, unified model that can be easily adapted for diverse downstream tasks and domains with minimal adjustments \cite{geng2022recommendation,cui2022m6}.

Recently, large language models (LLMs) have exhibited remarkable capabilities in semantic comprehension and knowledge inference, leading to impressive advancements in numerous downstream NLP tasks \cite{liu2022dial2vec,thoppilan2022lamda}. Recent studies have demonstrated that scaling up model size can endow LLMs with a range of generalization abilities, especially when provided with task descriptions or demonstrations \cite{qiao2022reasoning,brown2020language,wang2022self}. These capabilities provide promising opportunities to tackle existing challenges that require generalization for recommendation \cite{bao2023tallrec}.


Motivated by recent advancements in LLMs and prompt learning, LLM-based recommendation has emerged as a prominent research direction. In this approach, LLMs are acknowledged as foundational models, and recommendation tasks are reformulated as language generation through a unified sequence-to-sequence framework \cite{geng2022recommendation, cui2022m6, zhang2021language, wang2022towards, bao2023tallrec}. Previous studies treat user and item IDs as plain text, and directly input them into LLMs for recommendation tasks \cite{grbovic2015commerce, sun2019bert4rec, geng2022recommendation}. Although these methods have demonstrated significant progress, incorporating user and item IDs into LLMs presents challenges due to their unique semantic space, which differs from the natural language used during LLM pre-training \cite{li2023personalized}.

Another research approach suggests finding alternative plain text to represent IDs \cite{zhang2021language, cui2022m6, li2023personalized,hou2022towards,li2023text}. However, this conversion process may result in the loss of certain information \cite{li2023personalized}. To illustrate, let's consider a scenario where there are two online platforms selling a copy of the same movie, \textit{"The Avengers"}. Both platforms assign a unique item ID to differentiate the movie within their respective catalogs. By considering the unique IDs, the recommendation system can accurately recommend \textit{"The Avengers"} from a specific platform based on user preferences, pricing, streaming availability, or other relevant factors. Treating these two instances of \textit{"The Avengers"} as the same item without considering their unique IDs could lead to inaccurate recommendations or overlooking important nuances that influence the user's choice, such as the preferred streaming platform or subscription availability.

In order to tackle the aforementioned challenges, we introduce a novel \textbf{Contr}astive pr\textbf{o}mpt \textbf{l}earning framework for \textbf{Rec}ommendation System (ControlRec), which aims to bridge the semantic gap between natural language and IDs without compromising on accuracy of recommendation information. Inspired by the successful application of contrastive learning in multimodal learning \cite{wang2022towards,dong2022dreamartist}, we leverage similar principles to align and fuse textual information with IDs, creating a shared representation space that captures meaningful cross-modal correlations and semantic relationships. Specifically, the input of ControlRec is divided into two individual components, i.e., task-specific natural language (NL) instruction, termed as prompt, and a sequence of user and item IDs in the user interaction history. In contrast to previous approaches that concatenate these two parts and encode the resulting sequence as a whole, we treat them as heterogeneous features and encode them independently. To handle the semantic discrepancy problem when treating IDs as natural language, we introduce an auxiliary objective called Heterogeneous Feature Matching (HFM). HFM considers IDs and NL descriptions as two distinct views of the same item, and it optimizes an objective function that promotes consistency between positive pairs of views while maximizing the dissimilarity between negative pairs.



While HFM primarily focuses on aligning the representations of IDs and natural language on the encoder side, the main challenge lies in effectively fusing these two types of features on the decoder side. Although integrating these features in various recommendation tasks using the decoder partially alleviates this problem, it falls short in differentiating between the semantic meanings expressed by different NL instructions. This limitation often leads to model collapse and overfitting, particularly when the prompts are rare or absent during the training stage \cite{li2021align}. To this end, we present Instruction Contrastive Learning (ICL), which involves comparing the probability distributions of output sequences generated by different tasks. By employing this method, our model develops a deeper understanding of the relationships between prompts and IDs. Consequently, it enhances the model's ability to effectively merge the semantic meaning conveyed by IDs and natural language in recommendation systems.

For evaluation, we conduct comprehensive experiments on several benchmarks for each task and compare ControlRec with both task-specific and LLM-based recommendation systems. Experimental results demonstrate that ControlRec consistently outperforms other baselines on most metrics. Furthermore, we also conduct ablation studies to verify the design of each component. To summarize, our main contributions can be concluded as follows: 

\begin{itemize}
	\item We present a novel Contrastive Prompt Learning framework for LLM-based  Recommendation (ControlRec), which, to the best of our knowledge, is the first work that attempts to bridge the semantic discrepancy between language models and recommendations.
	\item We propose two auxiliary contrastive losses, namely Heterogeneous Feature Matching and Instruction Contrastive Learning, to align and merge IDs with natural language in semantic spaces on both the encoder and decoder sides respectively.
	\item We evaluate ControlRec on four public real-world recommendation datasets, and extensive quantitative experiments demonstrate the effectiveness of the proposed method.
\end{itemize}

\begin{figure*}[h]
	\centering
	\includegraphics[width=\linewidth]{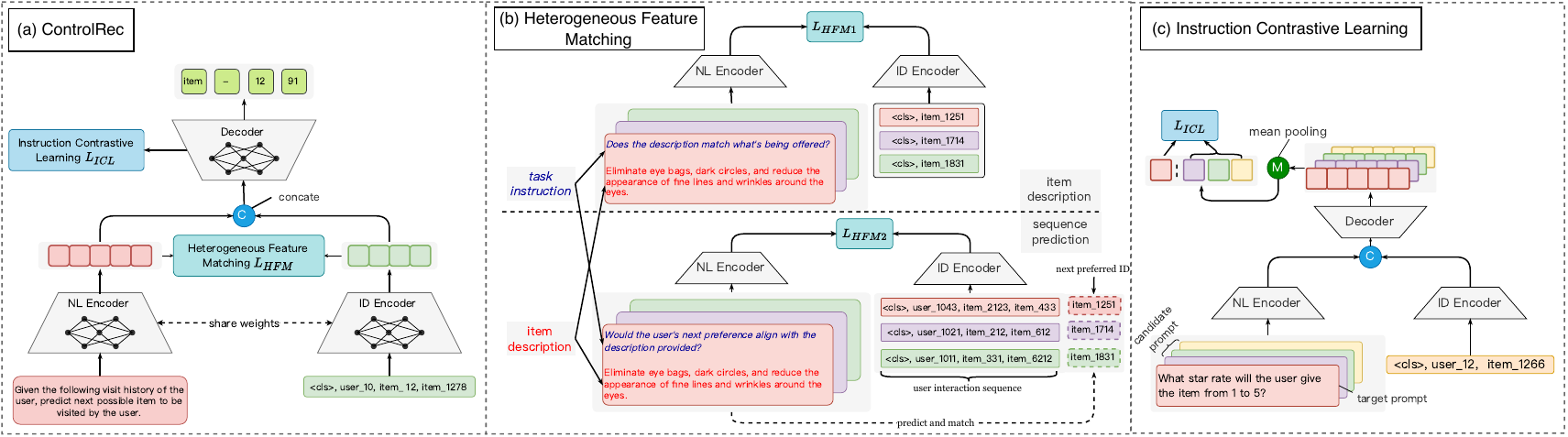}
	\caption{Illustration of the proposed ControlRec. }
	\label{fig:01}
\end{figure*}

\section{Related Work}
In this section, we will introduce existing works that are closely related to our research.

\textbf{LLM-based Recommendation Systems.}  With the widespread adoption of LLMs, the question of how to effectively adapt these models for recommendation systems has become a central focus for researchers in the recommendation community. According to the training paradigm, LLM-based recommendation systems generally fall into two categories. The first category follows the \textit{``pre-train and fine-tune''} paradigm. Under this category, these models incorporate language model (LM) based learning objectives during the pre-training phase and modify the holistic \cite{kang2021apirecx} or partial \cite{hou2022towards,yu2022tiny} model parameters, or even introduce additional components \cite{shang2019pre}, during the fine-tuning stage. Another approach in the research field follows the \textit{"pre-train, prompt, and predict"} paradigm, where the pre-trained model can directly perform downstream recommendation tasks guided by various prompts. This approach effectively overcomes the disparity in objective forms between the pre-training and fine-tuning stages \cite{liu2023pre}. Building upon this notion, our paper introduces the development of a unified LLM-based recommendation system that encompasses multiple tasks, aligning with the latter paradigm.

\textbf{Prompt Learning} refers to the process of prepending precise textual instructions \cite{brown2020language} or continuous embeddings \cite{liu2021gpt} to the task input, which improves the few-shot or zero-shot performance of LM across various downstream tasks. Instead of fine-tuning LM via objective engineering, it formulates downstream tasks to resemble objective of language modeling using prompts, which has already been addressed during the original LM training \cite{liu2021pre}. By providing a set of suitable prompts, a single language model that has been trained in an entirely unsupervised manner can be effectively utilized to perform a wide range of tasks. 

It has been demonstrated by previous pioneering studies that human-written instructional prompts are capable of eliciting knowledge from LLMs \cite{brown2020language}. Unfortunately, these methods typically require a significant amount of human effort. To alleviate the problem, several works leverage randomly initialized continuous embeddings, know as soft prompts, to replace discrete prompts. Nevertheless, soft prompts can be incompatible to reuse across models due to discrepancies in latent spaces \cite{deng2022rlprompt}. As an alternative, another line of research augments hand-crafted prompts with heuristics, such as paraphrasing-based methods for creating prompt templates introduced by LPAQA \cite{jiang2020can}. Similarly, other researchers have explored the modification of existing prompts using edit operations like deletion, swapping, and addition \cite{prasad2022grips}. However, the quality and quantity of generated prompts in these approaches can not be guaranteed. In light of these problems, we harness the powerful \textit{in-context learning} capabilities of LLMs \cite{brown2020language,ye2023context} to generate a substantial number of high-quality prompts, which promotes the robustness to the choice of prompts.

\textbf{Contrastive Learning (CL)} is a type of self-supervised learning (SSL) that has shown its potential in a wide range of application domains including natural language processing \cite{gao2021simcse,huang2022copner} and computer vision \cite{chen2020simple,dong2022dreamartist}. More recently, it has emerged as a popular and effective technique for generating high-quality representations of users and items, which in turn facilitates producing satisfactory recommendation results \cite{liu2021contrastive,wu2022userbert,xie2022contrastive}. The fundamental idea is to leverage data augmentation to produce diverse and augmented views of the same user interaction sequence, such as item dropout \cite{wu2022userbert}, cropping \cite{xie2022contrastive} and substitution \cite{liu2021contrastive}. These views are then employed to bring instances with comparable attributes closer together in the embedding space, while simultaneously pushing apart perspectives of distinct instances \cite{yu2022self}. For example, UserBERT \cite{wu2022userbert} masks part of items in the behavior sequence and pre-train the model in a contrastive way. CL4SRec \cite{xie2022contrastive} randomly selects a contiguous sub-sequence of the user behaviors to serve as an alternative to the original sequence. CoSeRec \cite{liu2021contrastive} replaces items in the interaction sequence with alternative items that are highly correlated with the originals, so as to minimize the amount of noise introduced to the original sequential information. Unlike previous studies, our research leverages CL to bridge the semantic gap between natural language and IDs. By doing so, we demonstrate significant improvements in the performance of LLM-based recommendation systems.

\section{Methodology}

Following P5 \cite{geng2022recommendation}, ControlRec is pre-trained and evaluated on five task families, including rating prediction, sequential recommendation, explanation generation, direct recommendation and review summarization. Likewise, we manually create a collection of prompt templates, referred to as \textit{triggers}, for each task family. These triggers are then fed into \textit{gpt-3.5-turbo}\footnote{\href{https://platform.openai.com/}{https://platform.openai.com/}} model to automatically generate more prompts. The details will be introduced in Section \ref{sec:ICL}.

We now introduce the proposed method from the following aspects: the \textbf{model architecture} and two contrastive modules: \textbf{Heterogeneous Feature Matching} and \textbf{Instruction Contrastive Learning}.


\subsection{Model Architecture}
In this section, we will provide an overview of our proposed ControlRec framework. As shown in Figure \ref{fig:01} (a), ControlRec, parameterized by $\mathbf{\Theta}$, comprises of an ID encoder $\mathcal{E}_{id}$, an NL encoder $\mathcal{E}_{nl}$, and a decoder $\mathcal{D}$. Notably, the weights between the two encoders are shared. The ID and NL inputs are encoded into a sequence of dense embeddings $\{\mathbf{h}^{id}_i\}_{i=0}^{\ell_{id}}$ and $\{\mathbf{h}^{nl}_j\}_{j=0}^{\ell_{nl}}$ respectively, where $\mathbf{h}_{0}$ is the output representation of <cls> token, $\ell_{id}$ and $\ell_{nl}$ represent the length of the input sequence. After obtaining the respective representations of both input sequences, we combine them to form a single input, which is then fed into the decoder for further generation. The training objective is modeled as predicting the probability distribution of future token $y_j$ based on the encoder-side vectors and previously generated words $y_{<j}$:
\begin{gather}
	\mathcal{L}_{gen}=-\sum_{j=1}^{|y|}\log P_{\mathbf{\Theta}}(y_j|y_{<j},[\mathbf{h}^{nl}_j,\mathbf{h}^{id}_j])
\end{gather}
Particularly, the ID input is a combination of a user ID and item IDs. While the user ID explicitly indicates the interaction subject, the form of item IDs can vary depending on the task at hand.
For rating prediction, explanation generation, and review summarization,  they refer to the item that is the main concern of the task. However, in the context of sequential and direct recommendation, they represent a sequence of IDs that denote the interaction history and the candidate items respectively. In traditional recommendation systems, each ID is considered as an independent identifier, which requires maintaining a massive embedding table if a large number of IDs are involved. Some research suggests the subdivision of words into multiple sub-words \cite{geng2022recommendation}. For instance, ``item\_1471'' which represents item with ID number 1471 will be tokenized as ``item'', ``\_'', ``14'' and ``71''. In this way, the IDs are treated as regular text and can be processed by LLMs. 

\begin{figure}[h]
	\centering
	\includegraphics[width=0.7\linewidth]{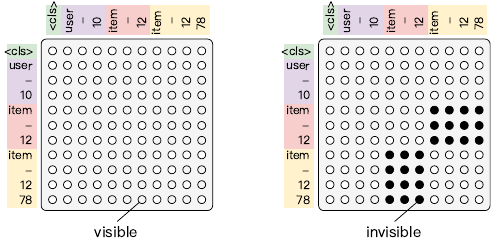}
	\caption{Visible matrix of traditional Transformer (left) and ours (right).}
	\label{fig:04}
\end{figure}

Nevertheless, Transformer lacks the ability to accurately model the interactions between user and items, which can be regarded as bipartite graph and is normally processed by graph neural networks (GNN) based models \cite{chen2020revisiting,he2020lightgcn,meng2022coarse}. Typically, two items do not directly interact with each other, but rather, their relationship can only be inferred through their interactions with the same user. However, traditional Transformer architecture allows tokens to attend to any other tokens within the same sequence. Thus we introduce a visible matrix into the conventional self-attention mechanism, which ensures that only relevant tokens are visible to each other during representation learning, as shown in Fig. \ref{fig:04}. 


\subsection{Heterogeneous Feature Matching}
After obtaining the representations of NL and IDs through their respective encoders, we introduce the Heterogeneous Feature Matching method to align the ID representations with the NL in the semantic space. This process improves the quality of the learned representations before combining them into the decoder for text generation. The primary objective is to bring the ID and NL features closer together in the semantic space when they have similar semantic meanings, while pushing them further apart when they are dissimilar. However, determining the similarity between two features still remains problematic.

Motivated by recent works that describe user interaction sequence with fine-grained item attributes or plain text behavior description \cite{li2023personalized,cui2022m6}, we assume that ID and NL features with similar semantic meaning should refer to the same target item. For instance, if the ID of a Nike t-shirt in an e-commence system is 7718, we say that the ID ``item\_7718'' matches  the description \textit{``Category: t-shirt. Brand: Nike''}, and we refer to them as a positive semantic pair. Based on this premise, we propose two sub-tasks called item description and sequential prediction, which can be seen in Fig. \ref{fig:01} (b).

\textbf{Item description} is designed to associate the item ID with its corresponding description, which could be an advertising commentary or a concatenation of its features, such as name, category, brand, or other relevant attributes. To achieve that, we propose to pre-train the model in a contrastive way. More concretely, we firstly sample $K$ descriptions randomly as well as a positive description to construct a candidate set $\mathcal{S}^{id}$ for each ID. To process the NL input for the task, we combine a task instruction (e.g., \textit{``Does the description match what's being offered?''}) with an item description. Then we employ a scoring function to calculate the semantic similarity between the ID and these $(K+1)$ NL candidates as follows:
\begin{gather}
	s(I,N)=\mathcal{E}_{id}(w_{cls})^T\cdot\mathcal{E}_{nl}(w_{cls})
\end{gather}
where $\mathcal{E}_{id}(w_{cls})$ and $\mathcal{E}_{nl}(w_{cls})$ denote the encoded output representation of token <cls> in the input sequence for ID and NL respectively. Similarly, for each description, we perform the same process to generate a candidate set that contains $K+1$ different IDs, and subsequently calculate the similarity scores between them. For each pair, ID-to-NL and NL-to-ID similarity scores are further normalized by softmax. The training objective is defined as the cross entropy between the normalized scores and the ground-truth one-hot similarity label $\mathbf{y}^{i2n}$ and $\mathbf{y}^{n2i}$ as follows:
\begin{gather}
	p^{i2n}_k=\dfrac{\exp(s(I,N_k)/\tau)}{\sum_{m=1}^{K+1}\exp(s(I,N_m)/\tau)} \label{eq:01} \\ 
	p^{n2i}_k=\dfrac{\exp(s(I_k,N)/\tau)}{\sum_{m=1}^{K+1}\exp(s(I_m,N)/\tau)} \label{eq:02} \\ 
	\mathcal{L}_{HFM1}=\dfrac{1}{K+1}\sum_{k=1}^{K+1}-(y^{i2n}_k\log p^{i2n}_k+y^{n2i}_k\log p^{n2i}_{k}) \label{eq:03}
\end{gather}
where $\tau$ is a temperature hyper-parameter, $y_k^{i2n}$ and $y_k^{n2i}$ equal to 1 for positive pairs and 0 for negative pairs.

\textbf{Sequential prediction} utilizes the interaction history to match the description of the user's next preferred item. To accomplish this task successfully, the model should be capable of conducting sequential recommendation, while also possessing the ability to match the predicted item IDs with their corresponding descriptions. In this task, we use the interaction history as ID input. The NL input resembles that of the item description task, but has customized task instructions tailored to prompt the model to perform sequential predictions (e.g., \textit{Would the user's next preference align with the description provided?}). Likewise, we utilize a scoring function to measure the similarity between user interaction history and item descriptions, along with similar sampling strategy to construct candidate sets. The training objective of sequential prediction $\mathcal{L}_{HFM2}$ is determined on Eq. (\ref{eq:01})$\sim$(\ref{eq:03}). The HFM loss is the combination of the two losses, i.e., $\mathcal{L}_{HFM}=\mathcal{L}_{HFM1}+\mathcal{L}_{HFM2}$.

\subsection{Instruction Contrastive Learning} \label{sec:ICL}
\begin{figure}[H]
	\centering
	\includegraphics[width=1\linewidth]{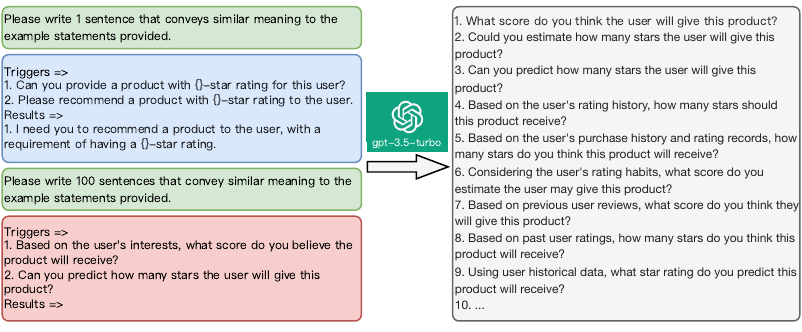}
	\caption{Prompt format for gpt-3.5-turbo (left), which consists of API instructions (green), demonstrations (blue) and contexts (red), and example of generated instructions (right).}
	\label{fig:05}
\end{figure}
Traditional prompt learning relies on a single, task-specific natural language instruction to guide the model to solve different tasks \cite{su2022multi,zhou2022learning,geng2022recommendation,cui2022m6}. However, this often causes model collapse and over-fitting when the prompts are rare or absent during the training stage \cite{dong2022dreamartist}. Especially, when performing recommendation tasks with identical ID inputs, the model may lack the ability to discriminate semantic differences between different task instructions. As a result, even minor changes of the expression can cause significant variation in the downstream task's performance \cite{liu2021gpt,xu2022making}. To alleviate this problem, we present instruction contrastive learning (ICL) on representations of the generated sequence. This approach not only enhances robustness of LLMs to the choice of prompts, but also facilitates the effective blending of the ID and NL inputs by enabling the model to differentiate between various prompts.

ICL pretrains LLMs on various prompt patterns in a contrastive way, which requires a substantial number of prompts. Therefore, data augmentation is crucial for ICL. P5 \cite{geng2022recommendation} manually creates a set of prompt templates for each task, which is time-consuming and incurs expensive human costs. Meanwhile, the limited number of collected prompts does not meet the requirements of ICL. Traditional techniques for automatic data augmentation in NLP include back translation\cite{sennrich2016improving}, synonym replacement \cite{zhang2015character} and sentence distortion \cite{wei2019eda}. However, these methods inevitably introduce noise to the synthetic data, which can have a detrimental effect on the quality and reliability of the generated prompts.

Fortunately, the emergence of AI-generative content (AIGC) presents promising solutions to produce abundant, high-quality prompts through \textit{In-context Learning} \cite{brown2020language}. In this paper, we utilize \textit{gpt-3.5-turbo}, the official OpenAI API, to generate task instructions. The prompt for \textit{gpt-3.5-turbo} is a concatenation of API instructions, demonstrations and contexts, as shown in Fig. \ref{fig:05}. Specifically, the API instructions are pre-set to generate a certain number of sentences that convey similar meaning as the given statements. The demonstrations consist of a fixed set of human-written statements, also known as \textit{"triggers"}, along with their corresponding results for few-shot in-context learning. The contexts resemble the demonstrations, but with the results left unfilled. Each task family is divided into several task groups to explore different aspects related to users and items\footnote{Manually designed prompts are not the primary focus of our research, and we do not delve into the details in this paper due to the space limitations. Readers can refer to P5 for more information.} \cite{geng2022recommendation}. 


Figure \ref{fig:04}(c) illustrates the ICL process, which involves performing various tasks using a target instruction and $(M+1)$ candidate instructions when given an ID input. These candidates are composed of one positive example that belongs to the same task group as the target prompt, along with $M$ negative examples selected from different task groups. ControlRec takes them as input and generates $(M+1)$ sequence representations $\{H_i \in \mathbb{R}^{\ell_i \times d_m}\}_{i=1}^{M+1}$, where $\ell_i$ is the length of the generated sequence and $d_m$ is the hidden size. To summarize the sequence representations into a fixed-size vector, we apply a mean-pooling operation on the generated sequences $\mathbf{u}_i=1/\ell_i\sum_{j=1}^{\ell_i}\mathbf{h}_{i,j}$, where $\mathbf{h}_{i,j}$ is the representation of the $j$th token in the generated sequence. Next, we compute the semantic similarity between the target and $(M+1)$ candidate representations using a scoring function to yield similarity scores, which are normalized by softmax. Then the ICL loss is defined as the cross entropy between the normalized scores $\mathbf{p}$ and ground-truth one-hot similarity label $\mathbf{y}^{ICL}$ as follows:
\begin{gather}
	p^{ICL}_i=\dfrac{\exp((\mathbf{u}_{0}^T\cdot\mathbf{u}_i)/\tau)}{\sum_{j=1}^{M+1}\exp((\mathbf{u}_{0}^T\cdot\mathbf{u}_i)/\tau)}\\
	\mathcal{L}_{ICL}=\dfrac{1}{M+1}\sum_{i=1}^{M+1}-y^{ICL}_i\log p^{ICL}_i
\end{gather}

where $y^{ICL}_i=1$ for positive candidate and $y^{ICL}_i=0$ for negative candidates. The final loss is the weighted sum of the aforementioned losses:
\begin{gather}
	\mathcal{L}=\lambda_1\mathcal{L}_{gen}+\lambda_2\mathcal{L}_{HFM}+\lambda_3\mathcal{L}_{ICL}
\end{gather}
where $\lambda_1,\lambda_2=1$. Given that ICL operates on the sequences generated by LLMs, it requires a strong generation capability from the model, which may be limited in the early training stage. Directly training the model with ICL may cause the learning process inefficient and unstable. Thus, we gradually increases $\lambda_3$ as the training processes. Supposing $T$ is the total number of training step, $\lambda_3'\in[0,1]$ is an initial value, $\lambda_3$ is calculated as follows:
\begin{gather}
	\lambda_3=\lambda_3'+\dfrac{1-\lambda_3'}{T}\cdot t
\end{gather}
In this paper, $\lambda_3'$ is set to $0$.

\section{Experiments}
In this section, we conduct experiments to assess the validity of our proposed ControlRec to answer the following research questions:\\
\textbf{RQ1:} How does the performance of ControlRec compare to that of task-specific and LLM-based recommendation systems?\\
\textbf{RQ2:} To what extent does each component of ControlRec contribute to its overall effectiveness?\\
\textbf{RQ3:} How does the number of generated prompts affect ControlRec's performance? \\
\textbf{RQ4:} How does the performance of ControlRec vary with different numbers of negative samples?\\

\subsection{Datasets}
Following P5 \cite{geng2022recommendation}, we conduct the experiments under the public Yelp\footnote{https://www.yelp.com/dataset} and Amazon\footnote{https://nijianmo.github.io/amazon/} datasets for a fair comparison. The Amazon dataset is a collection of data provided by Amazon, including information on 29 categories of products, such as product details, reviews, ratings, and other related information.  In this paper, we select three categories, including \textit{Sports \& Outdoors}, \textit{Beauty} and \textit{Toys \& Games} to evaluate the proposed ControlRec method. Yelp is a popular dataset for recommendation with a vast amount of user ratings and reviews.  As in original P5 paper \cite{geng2022recommendation}, We utilize transaction records from January 1, 2019 to December 31, 2019. 



\subsection{Implementation Details and Metrics}
In this paper, we employ T5-small and T5-base \cite{raffel2020exploring} models as the backbones to respectively initialize the weights of our proposed ControlRec \footnote{Models initialized with T5-small are denoted with a suffix "-S," while the models initialized with T5-based are denoted with a suffix "-B."}. The T5-small model consists of 60 million parameters, featuring 6 identical transformer layers. The model dimension is set to 512, which is divided into 8 attention heads. In contrast, the T5-base model is composed of 220 million parameters, with an embedding dimension of 768 and 12 attention heads. We utilize SentencePiece \cite{sennrich2015neural} with a vocabulary of 32000 wordpieces to tokenize words. The model is trained on 8 NVIDIA A100-80 GPUs for 10 epochs. The batch size for each GPU is set to 8 for T5-small and 4 for T5-base. It is trained using the AdamW optimizer with learning rate annealing from 1e-3 to 1e-6. The linear warm-up strategy is adopted at the first 5\% of all iterations. The number of negative samples in HFM and ICL, i.e., $K$ and $M$, is set to 10 and 5 respectively by default. We employ \textit{``gpt-3.5-turbo''} to generate 100 prompts for each task group, out of which we allocate 90 prompts for pre-training purposes. To evaluate zero-shot performance for each task, we randomly select 5 prompts that are not included in the training data. The remaining experimental settings are kept the same as in P5 to ensure a fair comparison.

Following P5 \cite{geng2022recommendation}, we evaluate ControlRec and other methods by adopting Root Mean Square Error (RMSE) and Mean Absolute Error (MAE) for rating prediction, top-$k$ Hit Ratio (HR@5/10 ) and Normalized Discounted Cumulative Gain (NDCG@5/10) for sequential and direct recommendations, BLEU-4 and ROUGE-1/2 for explanation generation and review summarization.

\subsection{Baselines}
We use the following 10 task-specific baselines for comparison:
\begin{itemize}
\item\textbf{MF} decomposes a user-item interaction matrix into lower-dimensional latent feature matrices, which capture the underlying characteristics of users and items, allowing the system to make accurate predictions.
\item\textbf{MLP} \cite{cheng2016wide} combines the strengths of wide linear models and deep neural networks for recommendation tasks.
\item \textbf{SASRec} \cite{kang2018self} strikes a balance between capturing long-term user behavior and predicting based on limited user actions in sequential recommendation tasks through a self-attention mechanism.
\item \textbf{S$^3$-Rec} \cite{zhou2020s3} utilizes the intrinsic data correlation to derive self-supervision signals and pre-training to enhance data representations for sequential recommendation.
\item \textbf{NRT} \cite{li2017neural} generates explanations based on user and item IDs using GRU \cite{chung2014empirical}.
\item \textbf{PETER} \cite{li2021personalized} designs a personalized transformer architecture to predict explanations conditioned on IDs.
\item \textbf{T0} \cite{sanhmultitask} and \textbf{GPT-2} \cite{radford2019language} are LLMs pretrained on massive natural language corpus. In P5 \cite{geng2022recommendation}, they are fine-tuned on recommendation tasks for review summarization.
\item \textbf{BPR-MF} \cite{rendle2012bpr} is a matrix factorization algorithm that utilizes Bayesian Personalized Ranking to optimize recommendation performance.
\item \textbf{SimpleX} \cite{mao2021simplex} incorporates contrastive learning into collaborative filtering model to facilitate recommendation.

\end{itemize}


Among them, MF and MLP are rating prediction baselines, SASRec, S$^3$-Rec are sequential recommendation baselines, NRT and PETER are explanation generation baselines, T0 and GPT-2 are review summarization baselines, BPR-MF and  SimpleX are direct recommendation baselines. We also compare our method with 2 LLM-based unified recommendation systems that are evaluated across all five tasks. 
\begin{itemize}
\item \textbf{P5} \cite{geng2022recommendation} leverages pre-training on LLMs and transforms recommendation tasks into customized natural language sentences using personalized prompts.
\item \textbf{LMRecSys} \cite{zhang2021language} reformulates recommendation as a language modeling task by converting a user’s interaction sequence to a text inquiry. In this paper, we implement LMRecSys by replacing item IDs with its attributes such as title, brand and categories.
\end{itemize}

\subsection{Experiment Results (RQ1)}
Table \ref{tab:rating} $\sim$ \ref{tab:direct} present the evaluation results on different task families. For each task, we evaluate the performance of LLM-based models by both providing them with prompts that are seen (\textcolor{blue}{S}) and zero-shot unseen (\textcolor{orange}{Z}) during training. In LMRecSys, the training and evaluation are performed using a single prompt for each task, following the same setting as P5 \cite{geng2022recommendation}. In the explanation generation and review summarization tasks, we notice that both P5 and LMRecSys use item titles instead of IDs to represent items. To avoid redundancy, we have decided not to present the results of LMRecSys alongside P5 in these two tasks. The first block of the table displays outcomes of task-specific models, the second block presents the results of LLM-based models and the last block showcases our proposed method\footnote{All results of task-specific and P5 models are borrowed from \cite{geng2022recommendation}.}. The best results are highlighted in \textbf{bold}, while the second-best results are \underline{underlined}.

\subsubsection{Rating Predicton}
Table \ref{tab:rating} presents the experimental results of rating prediction. It has been observed that, across all Amazon datasets, ControlRec-S outperforms other baseline models in terms of RMSE and MAE, and performs comparably on the Yelp dataset. Surprisingly, increasing the model size leads to overall performance degradation rather than improvement for ControlRec, which indicates the possibility of overfitting due to the task's low complexity. Nevertheless, ControlRec-B consistently surpasses other baseline approaches, demonstrating its ability to significantly enhance the performance of LLM-based models in rating prediction. Specifically, when compared with LMRecSys-S, ControlRec-S achieves notably higher performance on all evaluated datasets. These findings provide strong quantitative evidence for our claim that converting item IDs into natural language descriptions coupled with item attributes leads to a significant loss of information, which is fundamental to accurate recommendation.
\begin{table}[h]
\centering
\caption{Performance of rating prediction}
\label{tab:rating}
\resizebox{0.48\textwidth}{!}{%
	\begin{tabular}{@{}c|cc|cc|cc|cc@{}}
		\toprule
		\multirow{2}{*}{Methods} & \multicolumn{2}{c|}{\textbf{Sports}} & \multicolumn{2}{c|}{\textbf{Beauty}} & \multicolumn{2}{c|}{\textbf{Toys}} & \multicolumn{2}{c}{\textbf{Yelp}} \\ \cmidrule(l){2-9} 
		& RMSE & MAE & RMSE & MAE & RMSE & MAE & RMSE & MAE \\ \midrule
		MF & 1.0234 & 0.7935 & 1.1973 & 0.9461 & \textbf{1.0123} & 0.7984 & \textbf{1.2645} & 1.0426 \\
		MLP & 1.1277 & 0.7626 & 1.3078 & 0.9597 & 1.1215 & 0.8097 & 1.2951 & 1.0340 \\ \midrule
		P5-S (\textcolor{blue}{S}) & 1.0594 & 0.6639 & 1.3128 & 0.8428 & 1.0746 & 0.7054 & 1.4868 & 1.0186 \\
		P5-S (\textcolor{orange}{Z}) & 1.0522 & 0.6698 & 1.2989 & 0.8473 & 1.0550 & 0.7173 & 1.4685 & 1.0054 \\
		P5-B (\textcolor{blue}{S}) & 1.0357 & 0.6813 & 1.2843 & 0.8534 & 1.0544 & 0.7177 & - & - \\
		P5-B (\textcolor{orange}{Z}) & 1.0292 & 0.6864 & 1.2870 & 0.8531 & {1.0245} & 0.6931 & - &  -\\
		LMRecSys-S (\textcolor{blue}{S})  & 1.0491 & 0.7145 & 1.4152 & 0.9421 & 1.0912 & 0.7891 & 1.5613 & 1.3133 \\ 
		LMRecSys-S (\textcolor{orange}{Z}) & 1.0661 & 0.7912 & 1.4901 & 0.9821 & 1.1247 & 0.8013 & 1.4801 & 1.1401 \\ \midrule
		ControlRec-S (\textcolor{blue}{S}) & \textbf{0.9813} & \textbf{0.6081} & \underline{1.1820} & \underline{0.8191} & \underline{1.0178} & \textbf{0.6891} & 1.3528 &  {0.9781} \\ 
		ControlRec-S (\textcolor{orange}{Z}) & \underline{0.9878} & \underline{0.6371} & \textbf{1.1765} & \textbf{0.8142} & 1.0232 & \underline{0.6921} & 1.3458 & 0.9783 \\ 
		ControlRec-B (\textcolor{blue}{S}) & 1.0201 & 0.6451 & 1.2141 & {0.8217} & 1.0512 & 0.6978 & 1.2791 & \textbf{0.9689} \\ 
		ControlRec-B (\textcolor{orange}{Z}) & 1.0198 & 0.6451 & 1.2317 & 0.8243 & 1.0341 & 0.6973 & \underline{1.2785} & 0.9805 \\ 
		\bottomrule
	\end{tabular}%
}
\end{table}

\subsubsection{Sequential Recommendation}
Experimental results of sequential recommendation are illustrated in Table \ref{tab:sequential}. From the table, we can see that ControlRec-B achieves the best results on all datasets on both seen and unseen prompts. In contrast to rating prediction, we observe that larger model size boosts the performance by a considerable margin. Moreover, ControlRec demonstrates minimal gap between seen and unseen prompts, with state-of-the-art performance on Sports and Beauty datasets for seen prompts, and the best results on Toys and Yelp datasets for unseen prompts. We argue that it is attributed to the rich variety of prompts used during training.
\begin{table*}[h]
	\centering
	\caption{Performance of sequential recommendation (\%).}
	\label{tab:sequential}
		\begin{tabular}{@{}c|cc|cc|cc|cc@{}}
			\toprule
			\multirow{2}{*}{Methods} & \multicolumn{2}{c|}{\textbf{Sports}} & \multicolumn{2}{c|}{\textbf{Beauty}} & \multicolumn{2}{c|}{\textbf{Toys}} & \multicolumn{2}{c}{\textbf{Yelp}} \\ \cmidrule(l){2-9} 
			& HR@5/10 & NDCG@5/10 & HR@5/10 & NDCG@5/10 & HR@5/10 & NDCG@5/10 & HR@5/10 & NDCG@5/10 \\ \midrule
			SASRec & 2.33/3.50 & 1.54/1.92 & 3.87/6.05 & 2.49/3.18 & 4.63/6.75 & 3.06/3.74 & 1.62/2.74 & 1.00/1.36 \\
			S3-Rec & 2.51/3.85 & 1.61/2.04 & 3.87/6.47 & 2.44/3.27 & 4.43/7.00 & 2.94/3.76 & 2.01/3.41 & 1.23/1.68 \\ \midrule
			P5-S (\textcolor{blue}{S}) & 2.72/3.61 & 1.69/1.98 & 5.03/6.59 & 3.70/4.21 & 6.48/7.09 & 5.67/5.87 & 5.68/{7.07} & 4.02/4.47 \\
			P5-S (\textcolor{orange}{Z}) & 2.58/3.46 & 1.59/1.88 & 4.90/6.46 & 3.58/4.09 & \underline{6.48}/7.09 & 5.67/{5.87} & 5.74/7.03 & 4.03/4.45 \\
			P5-B (\textcolor{blue}{S}) & 3.64/4.31 & 2.96/3.18 & 5.08/6.64 & 3.79/4.29 & 6.08/6.88 & 5.07/5.34 & - & - \\
			P5-B (\textcolor{orange}{Z}) & 3.87/4.60 & 3.12/3.36 & 4.93/6.45 & 3.67/4.16 & 5.87/6.75 & 4.86/5.36 & - & - \\ 
			LMRecSys-S (\textcolor{blue}{S}) & 2.57/3.47 & 1.98/2.03 & 4.56/5.07 & 3.42/3.77 & 4.98/7.01 & 4.98/5.41 & 5.13/\underline{7.13} & 4.11/4.81 \\
			LMRecSys-S (\textcolor{orange}{Z}) & 2.45/3.41 & 1.83/2.01 & 4.53/5.01 & 2.85/3.21 & 3.81/5.57 & 4.11/4.89 & 4.56/6.16 & 3.54/4.03 \\\midrule
			ControlRec-S (\textcolor{blue}{S}) & 3.47/4.47 & 3.01/3.98 & 5.44/6.96 & 3.95/4.43 & 6.15/7.11 & 5.98/6.14 & 6.03/6.97 & 4.31/\underline{5.01} \\
			ControlRec-S (\textcolor{orange}{Z}) & 3.71/4.67 & 3.10/3.89 & 5.46/7.03 & 3.97/4.41 & 6.10/7.01 & 6.02/6.15 & 6.03/6.95 & 4.25/4.87 \\
			ControlRec-B (\textcolor{blue}{S}) & \textbf{4.31}/\textbf{5.11} & \textbf{3.56}/\textbf{4.71} & \textbf{5.69}/\textbf{7.81} & \textbf{4.31}/\textbf{4.97} &6.35/\underline{7.18} & \underline{6.49}/\underline{6.63} & \underline{6.11}/7.01 & \underline{4.57}/{4.97} \\ 
			ControlRec-B (\textcolor{orange}{Z}) & \underline{4.17}/\underline{5.02 }& \underline{3.41}/\underline{4.36} & \underline{5.47}/\underline{7.79} & \underline{4.27}/\underline{4.54} & \textbf{6.51}/\textbf{7.23} & \textbf{6.51}/\textbf{6.71} & \textbf{6.31}/\textbf{7.24} & \textbf{4.78}/\textbf{5.31} \\\bottomrule
		\end{tabular}%
\end{table*}

\subsubsection{Explanation Generation}
We now consider a more linguistic evaluation setting of studying a model’s generation performance, where it is asked to predict the explanation of a user towards an item.  In this experiment, we only consider the feature-based setup, i.e., the model will be provide with a topic related feature word to facilitate generation. The results are shown in Table \ref{tab:explanation}. In this task, P5 achieves competitive results, especially on Beauty and Toys datasets, which suggests that using purely natural language input has significant advantages in dealing with linguistically-related tasks. Nevertheless, ControlRec still achieves the best or comparable results on Amazon datasets, demonstrating the effectiveness of ControlRec in performing explanation generation via semantic alignment between IDs and natural language.
\begin{table*}[h]
	\centering
	\caption{Performance of explanation generation (\%).}
	\label{tab:explanation}
		\begin{tabular}{@{}c|cc|cc|cc|cc@{}}
			\toprule
			\multirow{2}{*}{Methods} & \multicolumn{2}{c|}{\textbf{Sports}} & \multicolumn{2}{c|}{\textbf{Beauty}} & \multicolumn{2}{c|}{\textbf{Toys}} & \multicolumn{2}{c}{\textbf{Yelp}} \\ \cmidrule(l){2-9} 
			& BLEU4 & ROUGE1/2 & BLEU4 & ROGUE1/2 & BLEU4 & ROUGE1/2 & BLEU4 & ROUGE1/2 \\ \midrule
			NRT & 0.47 & 11.07/1.13 & 0.82 & 12.78/1.85 & 1.90 & 13.52/3.67 & 0.81 & 13.92/1.96 \\
			PETER & \textbf{2.46} & 24.11/5.19 & \textbf{3.26} & 25.54/5.96 & {4.79} & {28.30}/9.45 & \textbf{3.28} & {27.23}/\textbf{8.19} \\ \midrule
			P5-S (\textcolor{blue}{S}) & 1.41 & 23.56/5.41 & 1.97 & 25.62/6.36 & 4.12 & {28.40}/9.54 & 2.97 & 27.18/6.68 \\
			P5-S (\textcolor{orange}{Z}) & 1.32 & 23.24/5.34 & 1.94 & 25.14/6.55 & {4.27} & 28.18/9.13 & \underline{3.06} & 27.29/6.76 \\
			P5-B (\textcolor{blue}{S}) & 1.46 & 23.54/5.39 & 1.97 & 25.62/6.36 & 4.12 & 27.99/{9.58} & - & - \\
			P5-B (\textcolor{orange}{Z}) &1.43  & 23.38/5.32 & 1.90 & 25.17/6.19 & 3.58 & 28.13/{9.75} & - & - \\
			\midrule
			ControlRec-S (\textcolor{blue}{S}) & 1.49 & 25.58/6.13 & 2.14 & 28.91/6.89 & 4.91 & 28.55/9.91 & 2.99 & 26.45/6.46 \\ 
			ControlRec-S (\textcolor{orange}{Z}) &{ 1.44} & {25.01}/6.13 & 2.17 & 29.14/ 6.98 & 4.99 & 28.91/9.65 & 2.67 & 26.31/6.65 \\ 
			ControlRec-B (\textcolor{blue}{S}) & \underline{1.52} & \underline{26.39}/\underline{6.48} & 2.38 & \underline{30.69}/\underline{7.31} & \underline{5.19} &\underline{28.97}/\underline{9.93} & {2.93} & \textbf{27.99}/6.79 \\ 
			ControlRec-B (\textcolor{orange}{Z}) & {1.50} & \textbf{26.78}/\textbf{6.65} & \underline{2.41} & \textbf{30.77}/\textbf{7.46}  & \textbf{5.21} & \textbf{28.98}/\textbf{10.01} & 2.95 & \underline{27.84}/\underline{6.89} \\ \bottomrule
		\end{tabular}%
\end{table*}

\subsubsection{Review Summarization}
Experimental results of review summarization are shown in Table \ref{tab:summarization}, ControlRec-S consistently outperforms all baseline models by a substantial margin. Despite that performance degradation is observed after increasing the model size, ControlRec-B still achieves markedly better results than P5. Overall, these findings confirm the effectiveness of our model in the review summarization task. 
\begin{table*}[h]
	\centering
	\caption{Performance of review summarization (\%).}
	\label{tab:summarization}
		\begin{tabular}{@{}c|cc|cc|cc@{}}
			\toprule
			\multirow{2}{*}{Methods} & \multicolumn{2}{c|}{\textbf{Sports}} & \multicolumn{2}{c|}{\textbf{Beauty}} & \multicolumn{2}{c}{\textbf{Toys}} \\ \cmidrule(l){2-7} 
			& BLEU2 & ROUGE1/2 & BLUE2 & ROUGE1/2 & BLUE2 & ROUGE1/2 \\ \midrule
			T0 & 2.15 & 2.26/0.56 & 1.28 & 1.27/0.39 & 2.22 & 2.46/0.64 \\
			GPT-2 & 0.77 & 4.45/1.00 & 0.58 & 3.38/0.67 & 0.62 & 3.71/0.66 \\ \midrule
			P5-S (\textcolor{blue}{S}) & 2.49 & 11.67/2.71 & 2.12 & 8.42/1.66 & 2.47 & 9.42/1.59 \\
			P5-B (\textcolor{blue}{S})  & 2.69 & 12.03/3.29 & 1.93 & 8.29/1.43 & 1.78 & 8.72/1.32\\	\midrule
			ControlRec-S (\textcolor{blue}{S})  & \textbf{3.14} & \underline{16.11}/\underline{4.14} & \underline{3.15} & \underline{10.55}/\underline{3.11} & \underline{2.91} & \underline{12.14}/\underline{1.91} \\ 
			ControlRec-S (\textcolor{orange}{Z}) & \underline{3.01} & \textbf{16.14/4.23} & \textbf{3.21} & \textbf{10.63/3.31} & \textbf{2.93} & \textbf{12.14/1.93} \\ 
			ControlRec-B (\textcolor{blue}{S}) & 2.89 & 14.79/3.99 & 2.96 & 9.37/2.17 & 2.60 & 10.57/1.78 \\ 
			ControlRec-B (\textcolor{orange}{Z})  & {2.91} & {14.89/4.01} & {2.98}& {9.41}/{2.19}& {2.61}& {10.69/1.78} \\ \bottomrule
		\end{tabular}%
\end{table*}

\subsubsection{Direct Recommendation}
The performance of direct recommendation is presented in Table \ref{tab:direct}. Our method achieves better results than BRP-MF, P5 and LMRecSys on all datasets. When comparing with SimpleX, ControlRec-B shows better NDCG performance on three Amazon datasets. In this task, increasing the model size boosts the performance significantly, with an average of 2\% HR and NDCG accuracy improvements. Moreover, LMRecSys falls behind the other two LLM-based models by a significant margin, which is similar to what is observed in sequential recommendation tasks. These observations indicate that representing items solely by their attributes is not enough to model collaborative filtering between users and items, which is crucial for accurate recommendation.

\begin{table*}[h] 
	\centering
	\caption{Performance of direct recommendation (\%).}
	\label{tab:direct}
		\begin{tabular}{@{}c|cc|cc|cc|cc@{}}
			\toprule
			\multirow{2}{*}{Methods} & \multicolumn{2}{c|}{\textbf{Sports}} & \multicolumn{2}{c|}{\textbf{Beauty}} & \multicolumn{2}{c|}{\textbf{Toys}} & \multicolumn{2}{c}{\textbf{Yelp}} \\ \cmidrule(l){2-9} 
			& HR@5/10 & NDCG@5/10 & HR@5/10 & NDCG@5/10 & HR@5/10 & NDCG@5/10 & HR@5/10 & NDCG@5/10 \\ \midrule
			BPR-MF & 14.04/25.63 & 8.48/12.20 & 14.26/25.73 & 8.57/12.24 & 10.66/20.03 & 6.41/9.40 & 22.51/33.12 & 15.43/18.86 \\
			SimpleX & \textbf{23.62/32.90} & 15.05/18.00 & \textbf{22.47}/\textbf{30.90} & {14.41/17.11}& \underline{19.58}/\textbf{26.62} & {12.44}/14.69 & \textbf{39.70/54.73} & \textbf{25.38/30.20 }\\ \midrule
			P5-S (\textcolor{blue}{S}) & 15.03/22.07 & 10.50/12.76 & 16.11/23.70 & 11.17/13.60 & 12.82/20.11 & 8.65/10.98 & 21.34/31.78 & 14.23/17.58 \\
			P5-S (\textcolor{orange}{Z})  & 15.14/21.96 & 10.49/12.69 & 15.66/23.17 & 10.78/13.18 & 13.22/20.23 & 8.89/11.14 & 21.36/32.19 & 14.32/17.80 \\
			P5-B (\textcolor{blue}{S}) & 17.94/25.98 & 12.29/14.88 & 15.73/23.25 & 10.89/13.30 & 11.22/18.07 & 7.56/9.75 & - & - \\
			P5-B (\textcolor{orange}{Z})  & 19.55/28.02 & 13.55/16.27 & 15.64/23.00 & 10.96/13.32 & 11.47/18.63 & 7.67/9.97 & - & - \\
			LMRecSys-S (\textcolor{blue}{S}) & 13.98/26.12 & 8.97/11.49 & 15.44/21.78 & 9.11/10.15 & 9.89/17.98 & 7.45/9.67 & 20.65/25.17 &  12.67/14.78 \\
			LMRecSys-S (\textcolor{orange}{Z})  & 13.01/23.45 & 8.15/10.23 & 14.76/19.77 & 8.91/9.01 & 9.55/16.66 & 7.34/9.11 & 19.78/25.66 & 12.67/14.34 \\ \midrule
			ControlRec-S (\textcolor{blue}{S}) & 20.11/26.12 & 12.41/14.56 & 18.13/25.56 & 15.41/16.09 & 16.16/23.47 & 10.10/12.09 & 24.11/34.97 & 16.78/22.45 \\
			ControlRec-S (\textcolor{orange}{Z}) & 20.31/26.77 & 12.56/15.01 & 18.23/25.57 & 15.51/16.13 & 16.32/23.46 & 10.23/12.91 & 25.14/35.17 & 16.99/23.15 \\
			ControlRec-B (\textcolor{blue}{S}) & 22.41/31.56 & \underline{15.10}/\underline{19.14} & {20.89}/{25.14} & \underline{17.02}/18.44 & 19.14/\underline{25.01} & \underline{12.77}/\underline{14.76} & \underline{26.98}/\underline{37.41} & \underline{18.41/}26.15 \\
			ControlRec-B (\textcolor{orange}{Z})  & \underline{22.58}/\underline{31.98} & \textbf{15.19/19.31} & \underline{20.98}/\underline{26.61} & \textbf{17.91}/\textbf{18.98 }& \textbf{19.98}/24.81 & \textbf{12.87}/\textbf{14.89}& 26.91/36.93 & 18.32/24.01 \\
			\bottomrule
		\end{tabular}%
\end{table*}

\subsection{Ablation Study (RQ2)}
In order to analyze the contribution of each component to the final performance, we conduct several ablation experiments by removing visible matrix (VM), HFM and ICL respectively on the Beauty dataset. For these experiments, we use the ControlRec-S model as the foundation and also provide results from  P5-S and LMRecSys-S for comparative analysis. All experiments are conducted in zero-shot setting to ensure fair evaluation across the models.

\begin{figure}[htbp]
	\centering
	\begin{subfigure}{0.22\textwidth}
		\includegraphics[width=\linewidth]{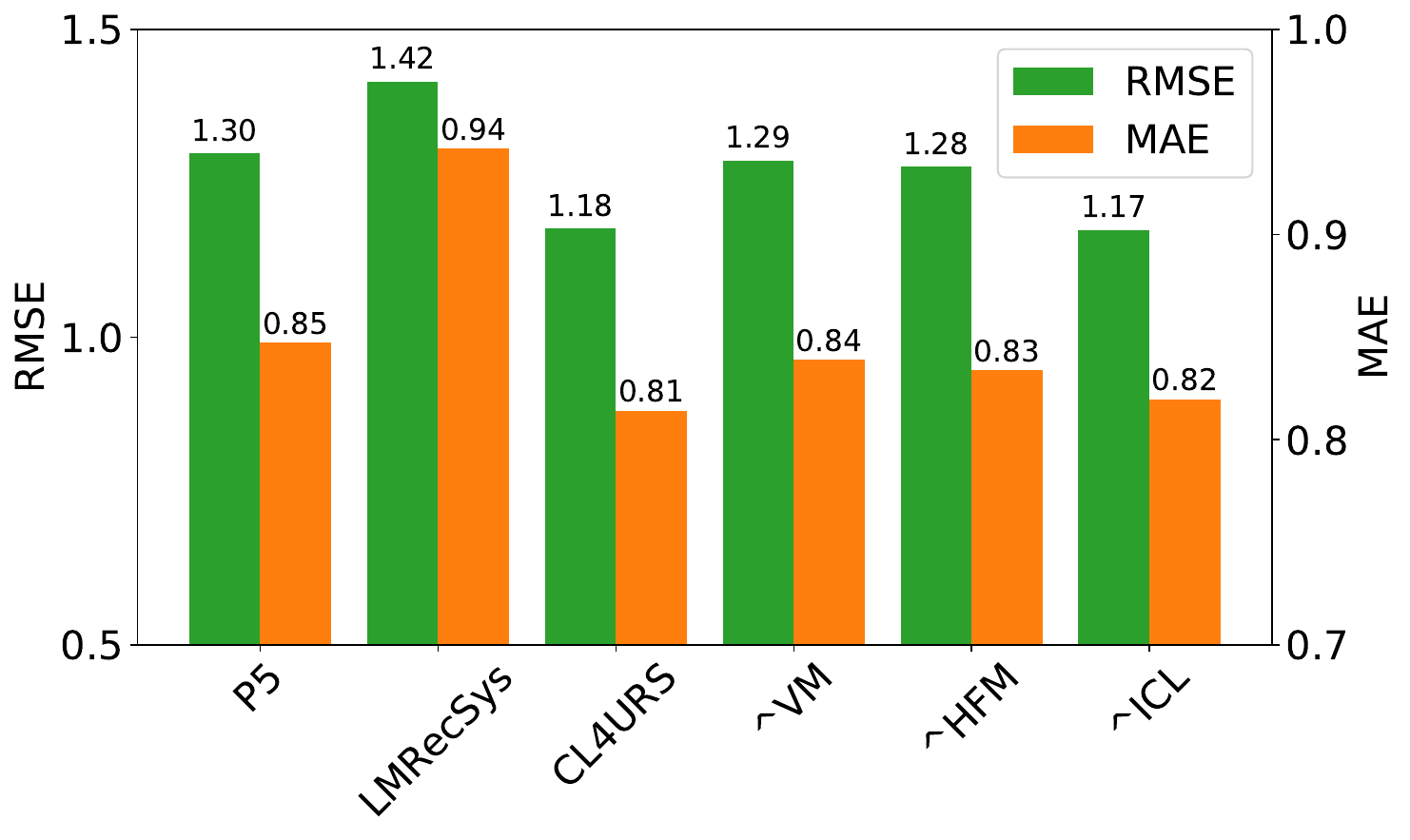}
		\caption{Rating Prediction}
		\label{fig:06a}
	\end{subfigure}
	\hfill
	\begin{subfigure}{0.22\textwidth}
		\includegraphics[width=\linewidth]{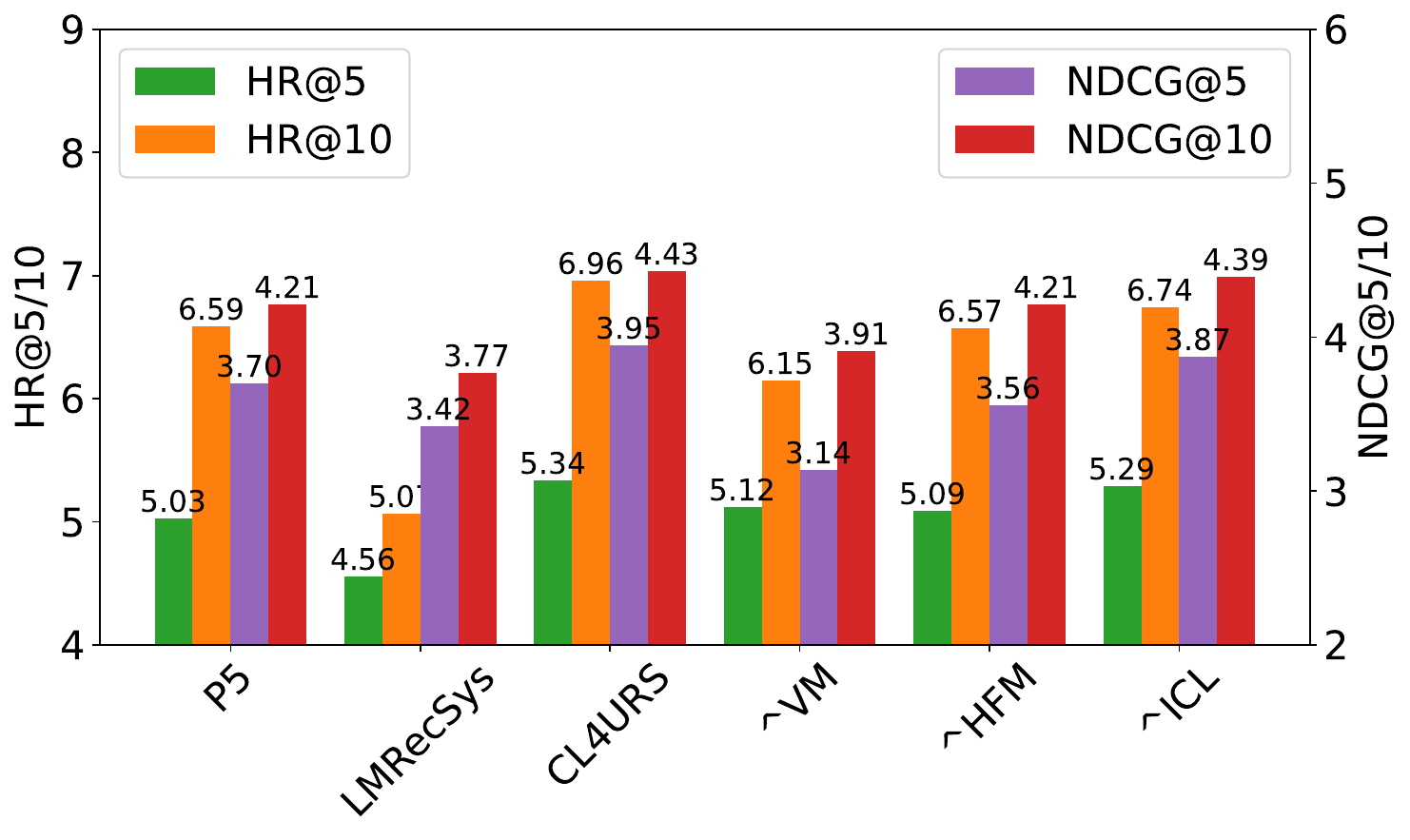}
		\caption{\footnotesize{Sequential Recommendation}}
		\label{fig:06b}
	\end{subfigure}
	\hfill
	\begin{subfigure}{0.22\textwidth}
		\includegraphics[width=\linewidth]{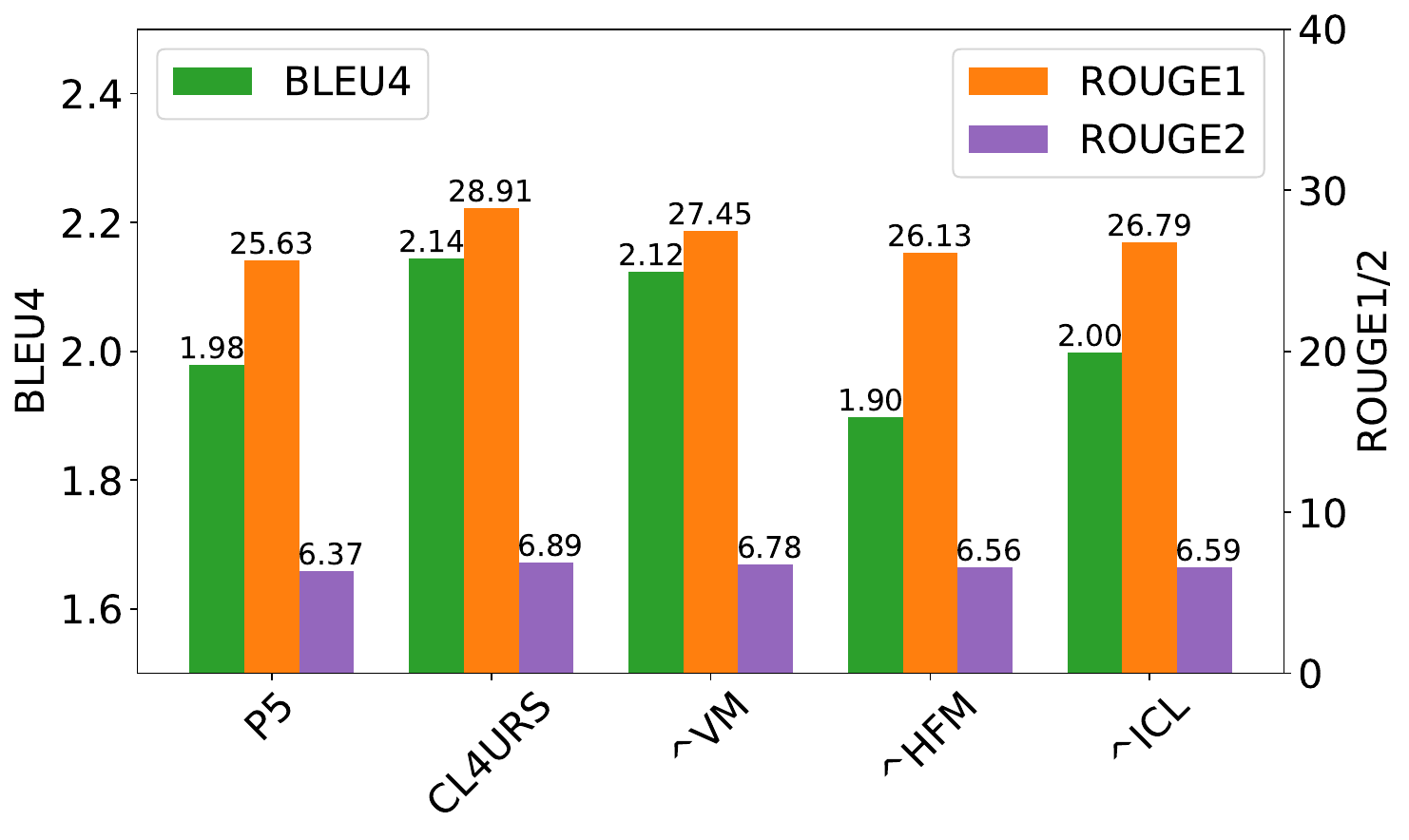}
		\caption{Explanation Generation}
		\label{fig:06c}
	\end{subfigure}
	\hfill
	\begin{subfigure}{0.22\textwidth}
		\includegraphics[width=\linewidth]{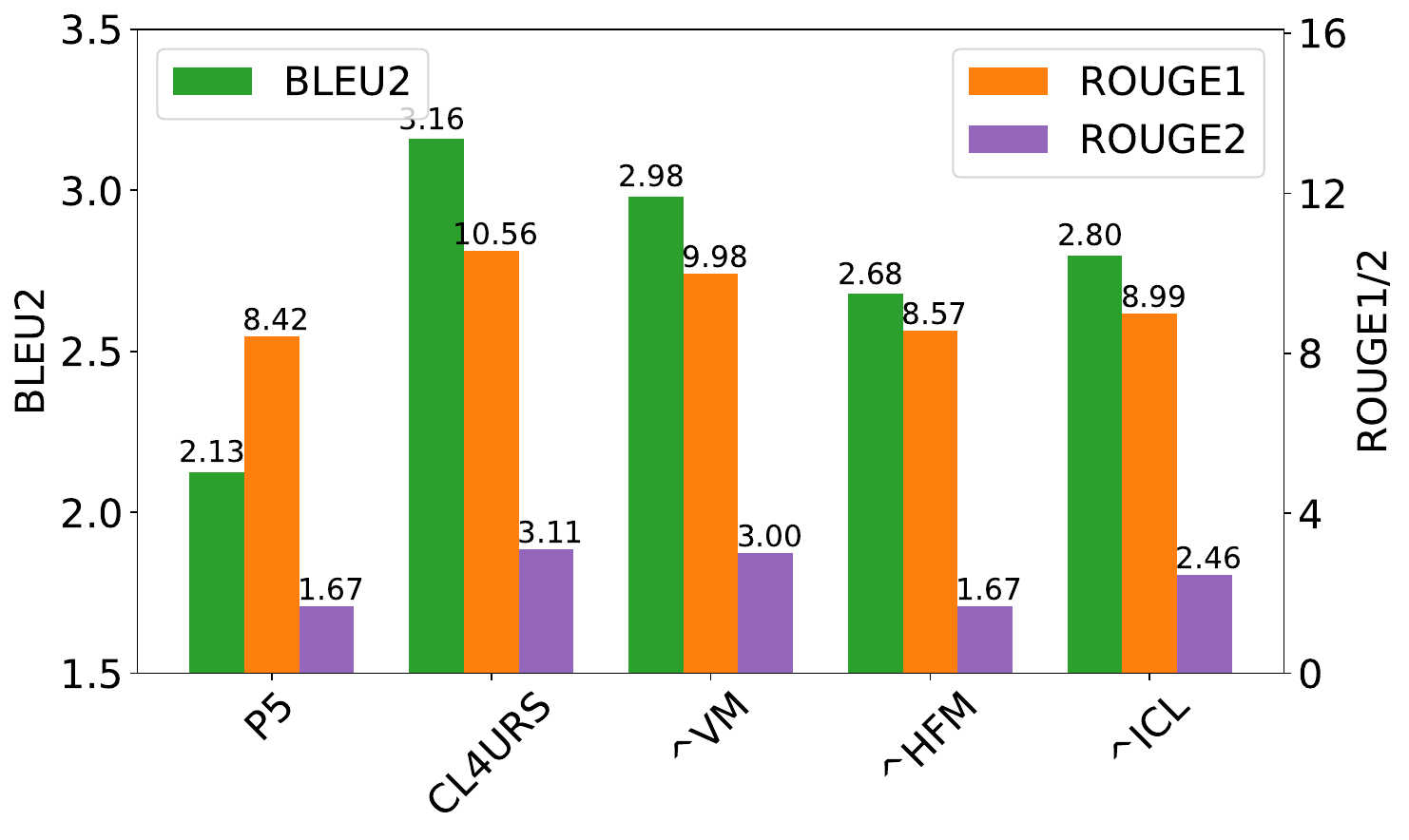}
		\caption{Review Summarization}
		\label{fig:06d}
	\end{subfigure}
	\hfill
	\begin{subfigure}{0.35\textwidth}
		\includegraphics[width=\linewidth]{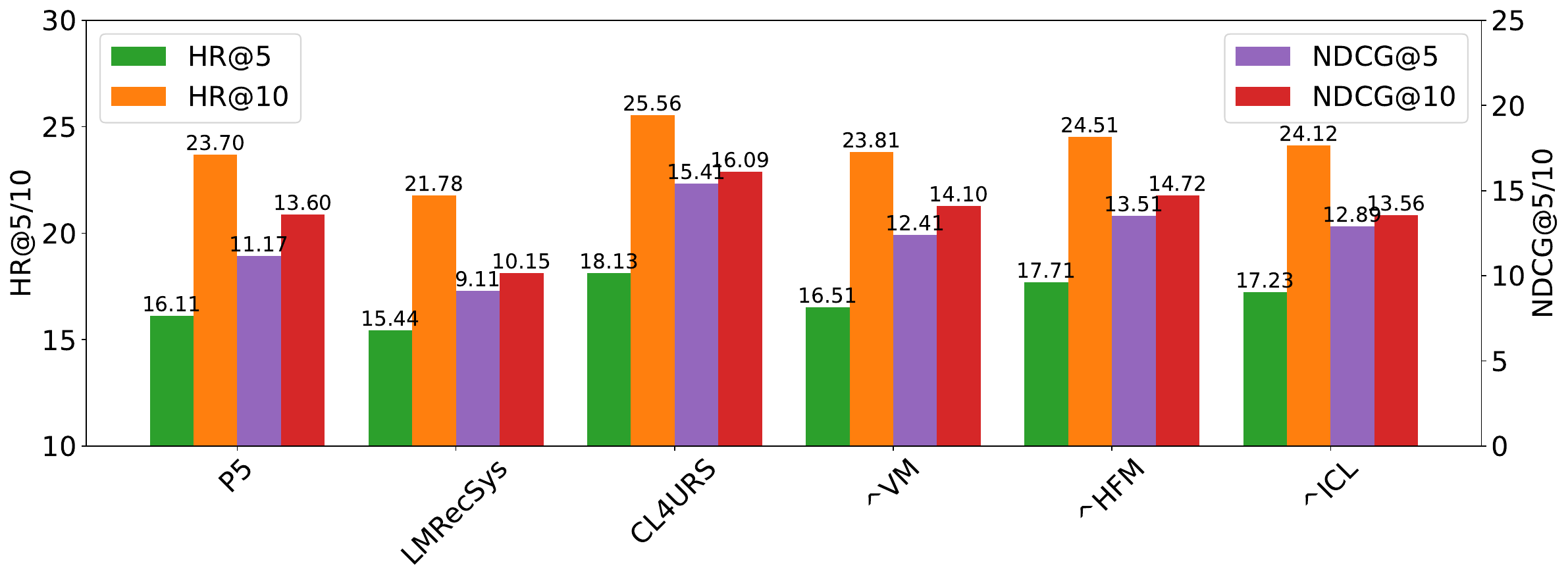}
		\caption{Direct Recommendation}
		\label{fig:06e}
	\end{subfigure}
	\caption[Ablation study of our method on Beauty dataset]{Ablation study of our method on Beauty dataset. ``\textasciicircum{}'' signifies the removal of corresponding modules during the pre-training stage, while preserving the rest of the modules.}
	\label{fig:06}
\end{figure}

The results, as depicted in Figure \ref{fig:06}, clearly demonstrate that the removal of any module results in performance degradation. Remarkably, almost all model variants outperform the LLM-based baselines (P5 and LLM-Recsys), emphasizing the crucial role played by each component in enhancing the overall recommendation performance. Additionally, it is noteworthy that the significance of these modules varies across different tasks. Particularly, VM exhibits greater importance in complex recommendation tasks like sequential and direct recommendation, as it effectively captures the intricate user-item interactions essential for accurate recommendations in such scenarios. Conversely, HFM and ICL demonstrate superior performance in simpler recommendation tasks, such as rating prediction, as well as linguistically-related tasks like explanation generation and review summarization. This phenomenon highlights the efficacy of the contrastive learning objectives in transforming IDs into the semantic space of natural language, allowing us to leverage the semantic reasoning and language generation capabilities of LLMs, thus maximizing their potential in recommendation systems.

\subsection{Performance Comparison w.r.t. the Amount of Prompts (RQ3)}
\begin{figure}[htbp]
	\centering
	\begin{subfigure}[b]{0.22\textwidth}
		\includegraphics[width=\textwidth]{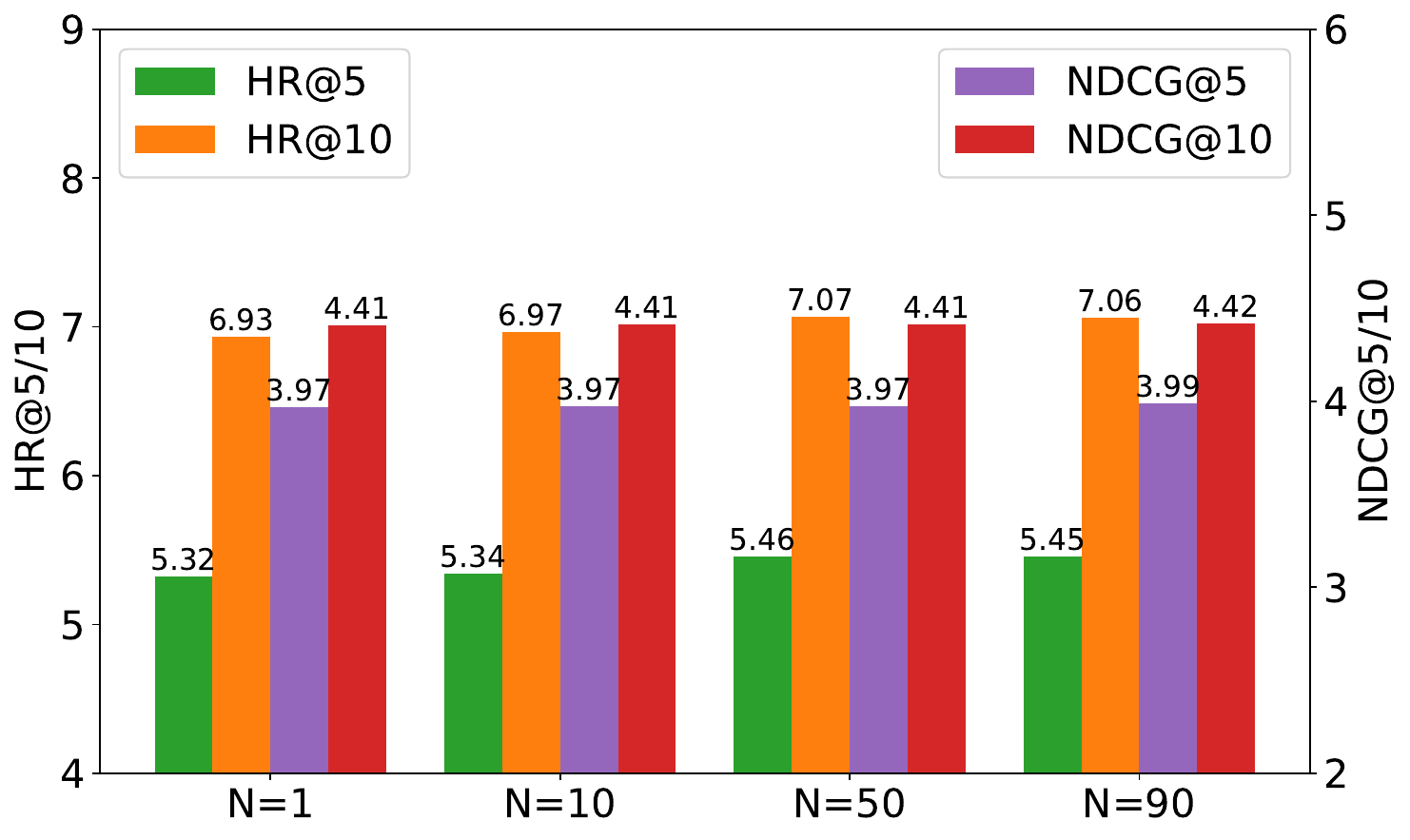}
		\caption{\footnotesize{Sequential Recommendation}}
		\label{fig:07a}
	\end{subfigure}
	\begin{subfigure}[b]{0.22\textwidth}
		\includegraphics[width=\textwidth]{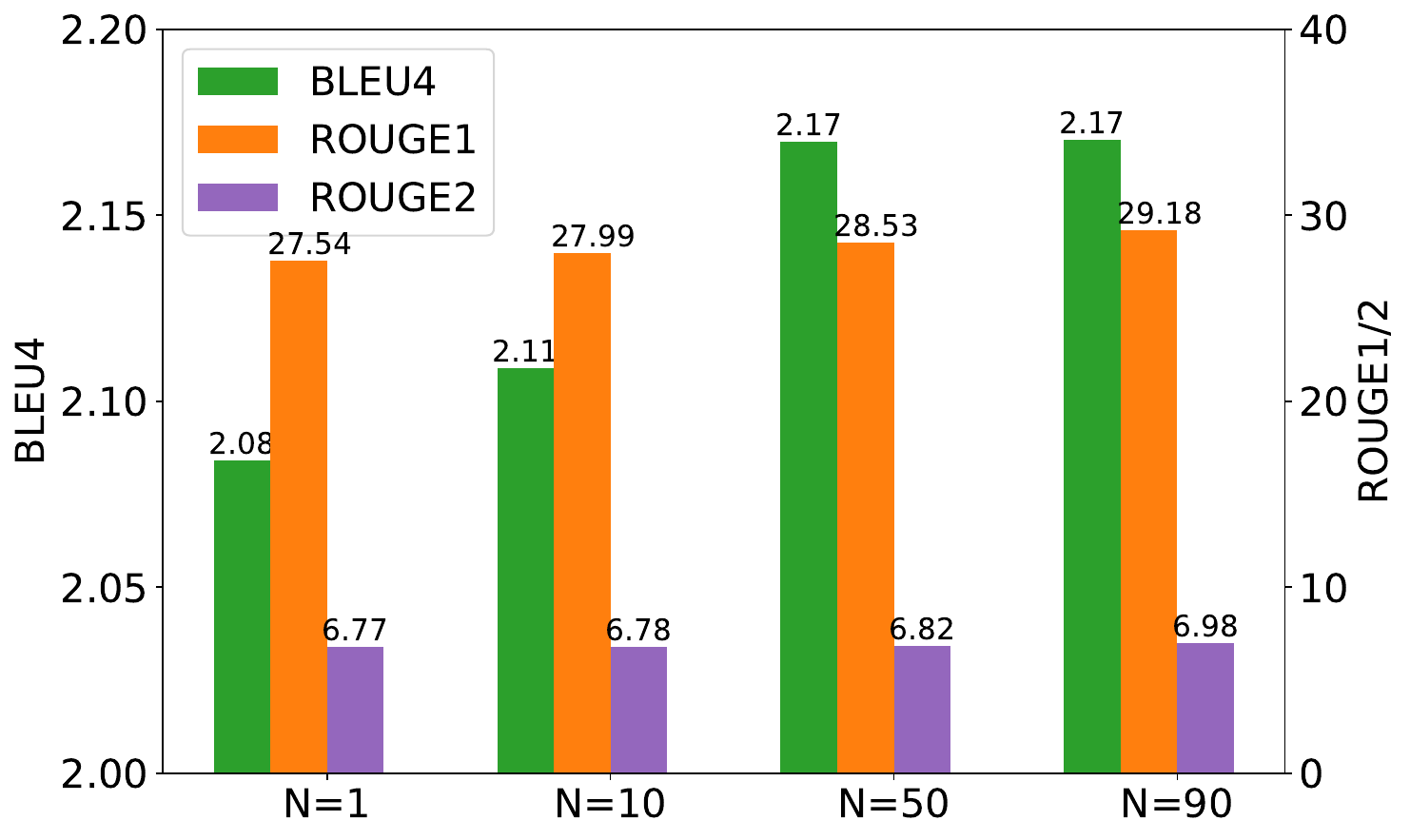}
		\caption{Explanation Generation}
		\label{fig:07b}
	\end{subfigure}
	\caption{Performance comparison w.r.t. different numbers of prompts on Beauty dataset.}
	\label{fig:07}
\end{figure}

To evaluate whether the number of prompts affects the zero-shot generalization ability of ControlRec, we pre-trained our model using different numbers of prompts (denoted as $N$) on the Beauty dataset. In these experiments, we randomly select $N$ out of 100 prompts generated by \textit{gpy-3.5-turbo} model for pre-training, and use 1 prompt selected from the rest of $(100-N)$ prompts for evaluation. The performance of sequential recommendation and review generation tasks are reported in Fig. \ref{fig:07}. As we can see, the performance shows a remarkable improvement as the number of prompts used for pre-training increases. This implies that the utilization of multiple prompts can significantly benefit the ControlRec model, leading to improved performance across various recommendation tasks in zero-shot setting. Besides, we notice that the review generation task yields a greater performance boost compared to the sequential recommendation task. This could be attributed to the fact that the review generation task requires a deeper understanding of language and context, thus benefiting more from the prompts augmentation during pre-training.

\begin{figure}[htbp]
	\centering
	\begin{subfigure}[b]{0.22\textwidth}
		\includegraphics[width=\textwidth]{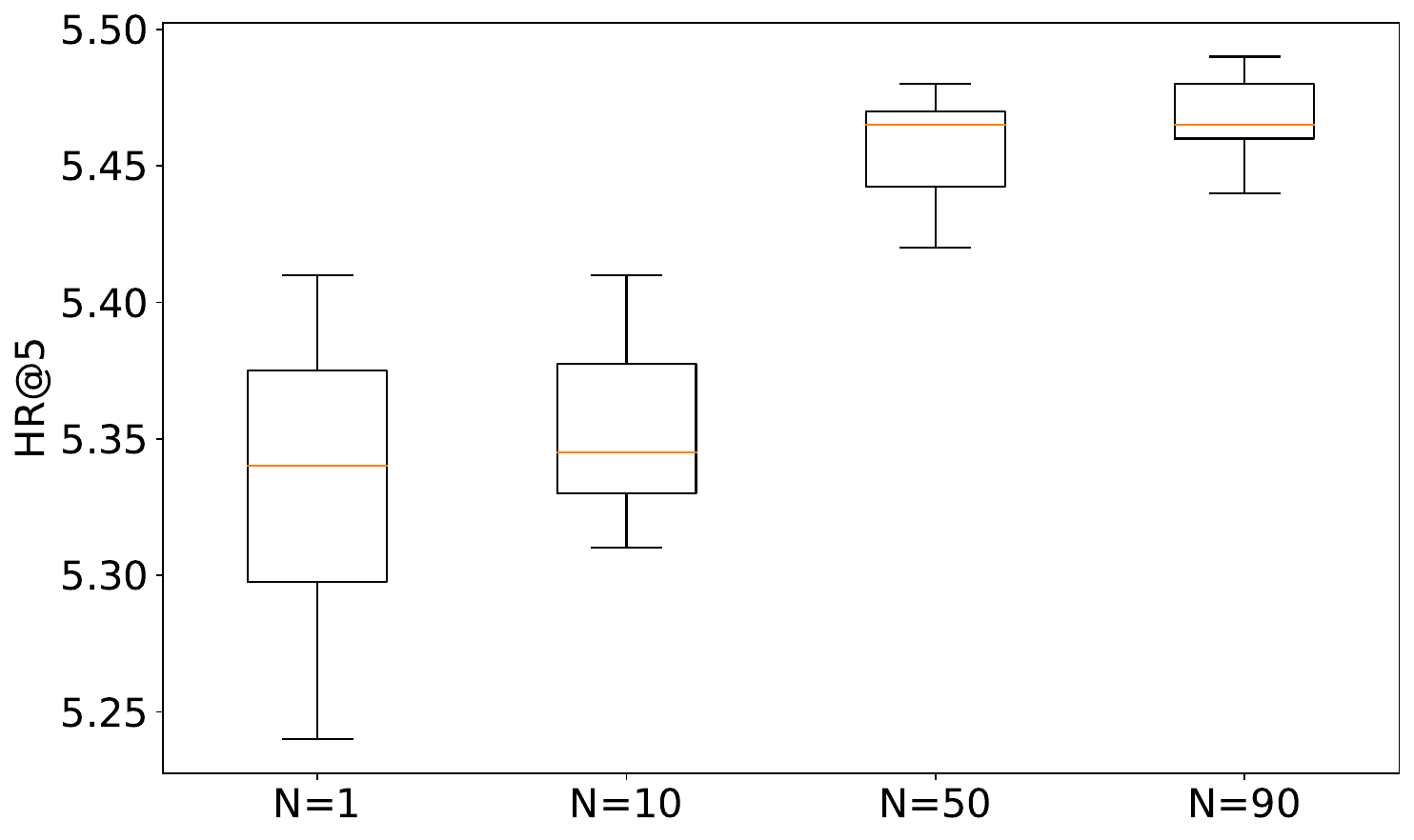}
		\caption{\footnotesize{Sequential Recommendation}}
		\label{fig:09a}
	\end{subfigure}
	\begin{subfigure}[b]{0.22\textwidth}
		\includegraphics[width=\textwidth]{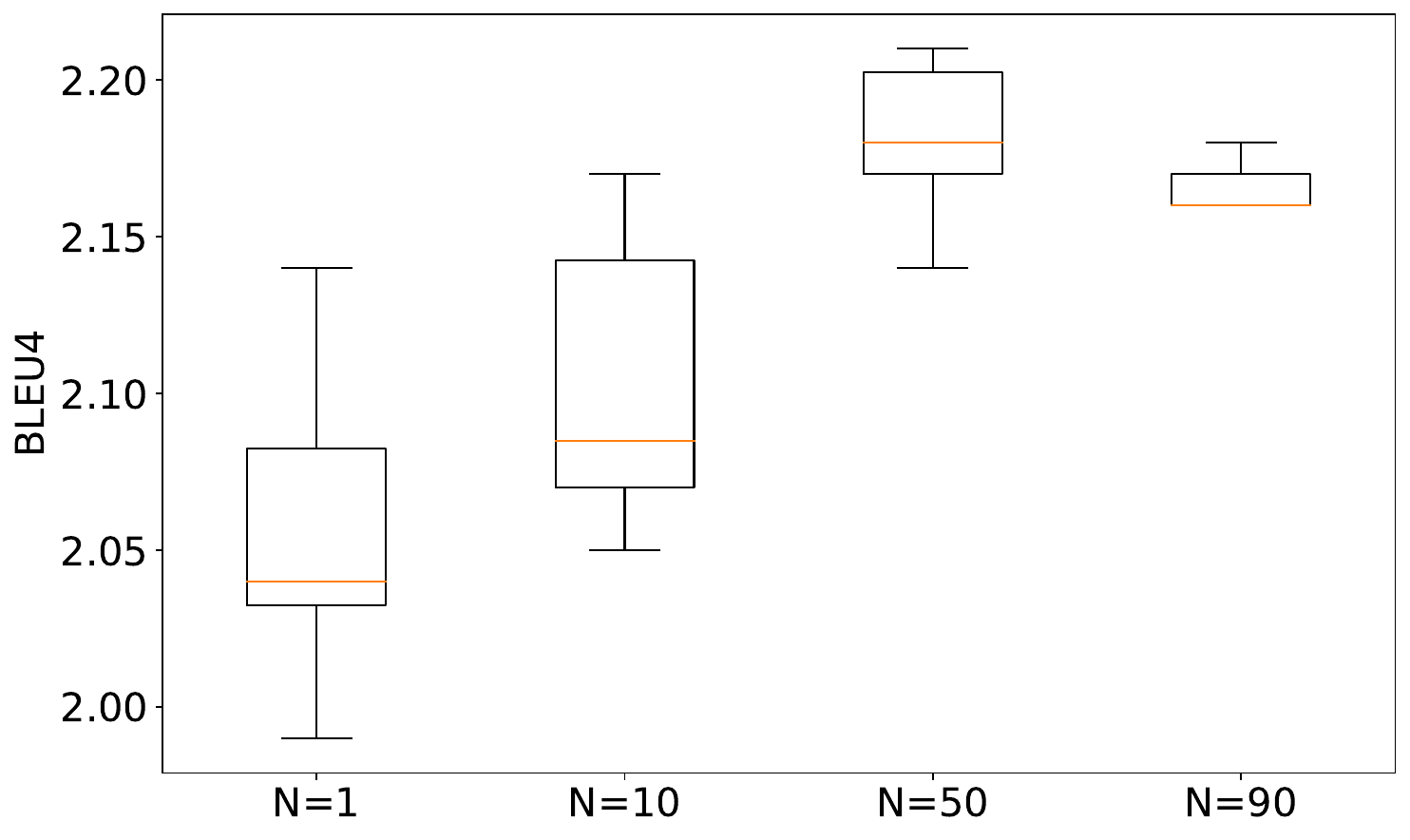}
		\caption{Explanation Generation}
		\label{fig:09b}
	\end{subfigure}
	\caption{Box plots w.r.t. different numbers of prompts on Beauty dataset.}
	\label{fig:09}
\end{figure}

Fig. \ref{fig:09} depicts the box plots showcasing the performance stability of our model with varying numbers of prompts. Each box plot represents 10 repeated evaluations, where the model is presented with a distinct prompt that was not encountered during pre-training. The results clearly demonstrate that models trained with a greater number of prompts exhibit enhanced robustness to the selection of prompts compared to those trained with fewer prompts. The findings indicate that ControlRec, leveraging the advantage of multiple prompts, achieves a more effective and stable performance.

\subsection{Performance Comparison w.r.t. the Amount of Negative Samples (RQ4)}
\begin{figure}[htbp]
	\centering
	\begin{subfigure}[b]{0.22\textwidth}
		\includegraphics[width=\textwidth]{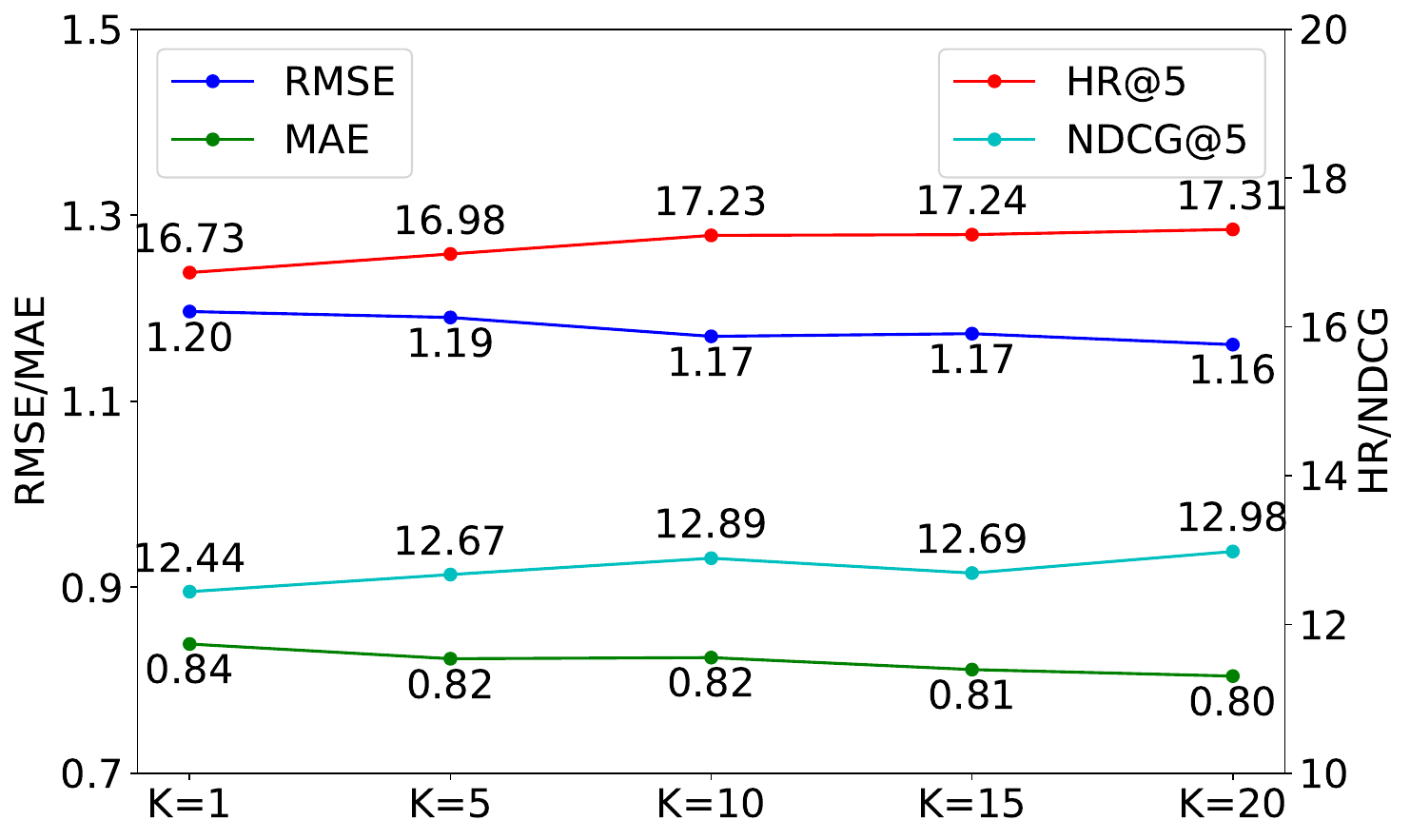}
		\caption{HFM}
		\label{fig:08a}
	\end{subfigure}
	\begin{subfigure}[b]{0.22\textwidth}
		\includegraphics[width=\textwidth]{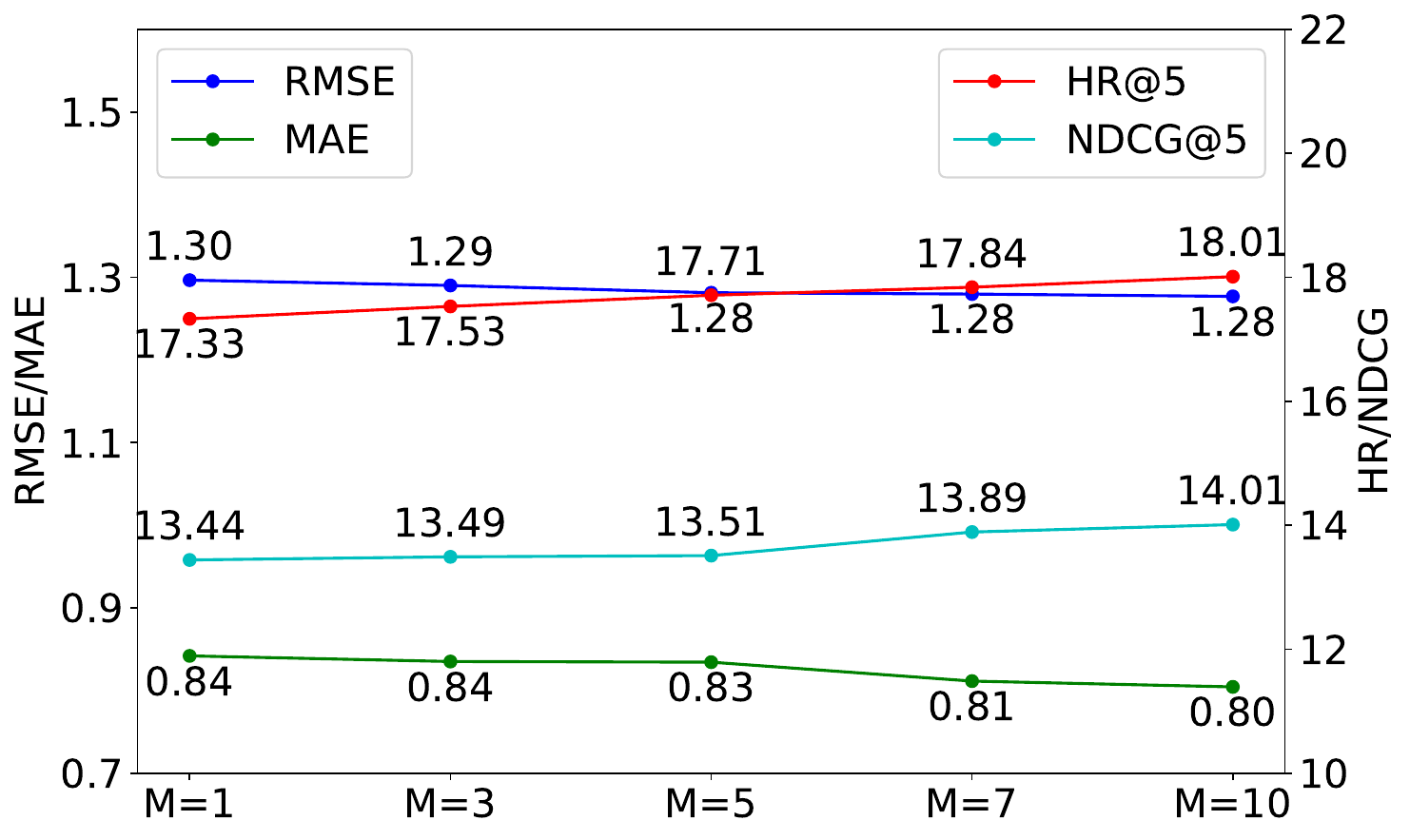}
		\caption{ICL}
		\label{fig:08b}
	\end{subfigure}
	\caption{Performance comparison w.r.t. different numbers of negative prompts on Beauty dataset.}
	\label{fig:08}
\end{figure}

In this section, we delve into investigating the influence of the number of negative samples in HFM ($K$) and ICL ($M$). The adopted foundation model is ControlRec-S (\textcolor{orange}{Z}). To isolate the impact of negative samples on HFM, we eliminate the ICL component during evaluation. Similarly, when assessing ICL, HFM is excluded from the analysis. We conduct experiments on the Beauty dataset and present the results for rating prediction and direct recommendation in Figure \ref{fig:08}. It is evident that the performance consistently improves as the number of negative samples increases, clearly demonstrating the positive impact of a larger number of negative samples on recommendation tasks. However, considering the additional time and computational resources required for pre-training the model with more negative samples, we strike a balance between performance and efficiency by setting $K=10$ and $M=5$ respectively.

\section{Conclusion and Future Work}
In this paper, we introduce ControlRec, which narrows the semantic gap between language models and recommendation systems via two auxiliary contrastive objectives, namely Heterogeneous Feature Matching (HFM) and Instruction Contrastive Learning (ICL). These objectives serve two main purposes: firstly, to align the natural language and ID representations, thereby facilitating language modeling for LLMs; and secondly, to teach the model to distinguish semantic differences when performing various tasks, thus improving its ability to integrate the two types of data sources. Extensive experiments and ablation studies validate the effectiveness of our proposed method.


However, our method, which relies on IDs to perform recommendation tasks, encounters limitations when dealing with cold-start recommendations. In the future, we will address this problem by developing new indexing techniques that assign IDs to items based on their correlations, rather than using random indexing. Furthermore, the training speed is significantly slowed down by conducting contrastive learning, especially with ICL, where the model operates in an autoregressive paradigm. We will also focus on addressing this issue in our future work.

\newpage
\bibliographystyle{ACM-Reference-Format}
\bibliography{acmart.bib}

\end{document}